\documentclass[a4paper,twocolumn,11pt,accepted=2024-12-15]{quantumarticle}
\pdfoutput=1
\usepackage[utf8]{inputenc}
\usepackage[english]{babel}
\usepackage[T1]{fontenc}
\usepackage[numbers,compress]{natbib}
\usepackage{amssymb, amsfonts, amsmath, amsthm, amsthm, bbm, braket}
\usepackage[final]{changes}
\usepackage{hyperref}
\usepackage{lipsum}
\def\unity{\mathbbm 1}
\def\tr{\mathrm{tr}}

\def\SU{\mathrm{SU}}
\def\A{\mathsf A}
\def\B{\mathsf B}

\def\F{\mathcal F}
\def\G{\mathcal G}

\def\u{\mathbf u}
\def\f{\mathbf f}
\def\g{\mathbf g}
\def\complex{\mathbb C}
\def\real{\mathbb R}
\newcommand{\proj}[1]{|#1\rangle\! \langle#1|}

\begin{document}

\title{Response to ``The measurement postulates of quantum mechanics are not redundant''}

\author{Llu\'\i s Masanes}
\email{l.masanes@ucl.ac.uk}
\affiliation{London Centre for Nanotechnology and Department of Computer Science, University College London, United Kingdom}
\author{Thomas D.~Galley} 
%\homepage{}
%\orcid{}
\affiliation{Institute for Quantum Optics and Quantum Information, Austrian Academy of Sciences, Boltzmanngasse 3, A-1090 Vienna, Austria}
\author{Markus P.~M\"uller}
\affiliation{Institute for Quantum Optics and Quantum Information, Austrian Academy of Sciences, Boltzmanngasse 3, A-1090 Vienna, Austria}
\affiliation{Perimeter Institute for Theoretical Physics, Waterloo, ON N2L 2Y5, Canada}

\begin{abstract}
Adrian Kent has recently presented a critique \cite{Kent} of our paper \cite{MGM} in which he claims to refute our main result: the measurement postulates of quantum mechanics can be derived from the rest of postulates, once we assume that the set of mixed states of a finite-dimensional Hilbert space is finite-dimensional.
To construct his argument, Kent considers theories resulting from supplementing quantum mechanics with hypothetical ``post-quantum'' measurement devices.
We prove that each of these theories contains pure states (i.e.~states of maximal knowledge) which are not rays of the Hilbert space, in contradiction with the ``pure state postulate'' of quantum mechanics. 
We also prove that these alternatives violate the finite-dimensionality of mixed states.
Each of these two facts separately invalidates the refutation.
In this note we also clarify the assumptions used in \cite{MGM} and discuss the notions of pure state, physical system, and the sensitivity of the structure of the state space under modifications of the measurements or the dynamics.
\end{abstract}

\maketitle

\medskip
The ``operational framework'' used in the derivation of the measurement postulates in \cite{MGM} is based on the operations that an experimentalist can ideally perform in a lab; and considers processes like preparations, transformations and measurements as primitive, without necessarily implying that they are fundamental.
Consequently, we assume that the composition and mixture of valid processes produces valid processes again.
Also, we require that physical systems have a well-defined state irrespective of which operations are performed on other systems, which implies the impossibility of signalling by measurements (sometimes referred to as ``no signalling'').
In Section~\ref{sec:assumptions} we detail all the assumptions made in \cite{MGM} and prove that no signalling follows from them.

The operational framework sketched above constitutes the bare bones of general probabilistic theories or GPTs \cite{Hardy_quantum_2001, Barrett_information_2005, Dakic_quantum_2009, Masanes_derivation_2011}. However, many assumptions often made in GPTs (e.g.~``local tomography'' \cite{Hardy_quantum_2001, Barrett_information_2005}) are not required here.
Instead, we assume the non-measurement postulates of quantum mechanics. In particular, the ``pure states postulate'' restricts state spaces to those whose pure states are the rays $\psi\in {\rm P}\complex^d$ of a Hilbert space $\complex^d$.
This still leaves a lot of freedom for the structure of mixed states, but we assume the following condition: the set of mixed states associated to a finite-dimensional Hilbert space $\complex^d$ is finite-dimensional too.
We agree with Kent \cite{Kent} in that \added{this condition is sufficient but not necessary for the possibility of state estimation.} However, this condition is meaningful and much weaker than stating the structure of quantum measurements and the Born Rule.

There exist alternative frameworks to the one outlined above. In~\cite{Kent}, Kent introduces new theories by supplementing quantum mechanics with hypothetical post-quantum measurements, and claims that these theories provide counter-examples to the statement that only the quantum measurement postulates are compatible with the rest of postulates of quantum mechanics. 
These new theories do not completely fit in our framework because they allow for signalling by measurements.
However, since this discussion is about our results in \cite{MGM}, we need to analyse them within our framework. We also think that this analysis is very fruitful, since it reveals features of Kent's theories that are not apparent at first sight.

The first feature is that these new theories have pure states (i.e.~extreme points of the state space)  other than the rays of the Hilbert space. Specifically, in some of these theories, the set of pure states of system $\complex^d$ is the corresponding set of $d\! \times\! d$ density matrices.

The second feature follows from the fact that, if we supplement quantum mechanics with a post-quantum measurement $\mathsf M$, then the resulting theory must also contain the measurement consisting of $\mathsf M$ preceded by any quantum dynamics. For some choices of $\mathsf M$, this proliferation of measurements produces an infinite-dimensional space of effects, which implies that the corresponding set of mixed states is also infinite-dimensional, in contradiction with the above-mentioned assumption.

Other features of the theories introduced in \cite{Kent} are detailed below. However, each of the above two, separately, already invalidates Kent's refutation \cite{Kent}.
At a mathematical level, we have no evidence of any mistake in the proof of our claim in \cite{MGM}, hence, the discussion should focus on the validity and meaning of our assumptions.

In Section~\ref{sec:GPT} we provide a quick introduction to the operational framework, in Section~\ref{sec:Kent theories} we analyse Kent's new theories, and in Section~\ref{sec:assumptions} we detail all the assumptions used for the derivation of the measurement postulates in \cite{MGM}. 

\section{Quick introduction to the operational framework}

\noindent
More detailed introductions can be found in \cite{Hardy_quantum_2001, Barrett_information_2005} and Appendix B of \cite{Masanes_existence_2013}.

\subsection{States and measurements}\label{sec:GPT}

\noindent
Consider a physical system $\A$ and denote by $\F_\A$ the set of all outcomes of all measurements that can be performed on $\A$ in principle, \added{according to some possible idealised physical theory}.
Analogously, consider the set of all preparations $\Lambda_\A$ of $\A$.
The probability of outcome $\f \in \F_\A$ on preparation $\lambda \in \Lambda_A$ is denoted by $P(\f|\lambda)$.

Let $\G_\A =\{\g_1, \g_2, \ldots \} \subseteq \F_\A$ be the smallest set of outcomes such that knowing their probabilities $P(\g_i|\lambda)$ on any preparation $\lambda \in \Lambda_A$ allows one to predict the probabilities $P(\f|\lambda)$ of any outcome $\f\in \F_\A$ on preparation $\lambda$.
This set of fiducial outcomes $\G_\A$ can be finite or countably infinite.
All possible predictions on the observable behaviour (including dynamics, measurements, etc) of preparation $\lambda\in \Lambda_\A$ are contained in the vector 
\begin{align}\label{omega vector}
  \omega_\lambda =
  \Big[P(\g_1|\lambda), P(\g_2|\lambda), \ldots\Big] \in \Omega_\A .
\end{align}
If this were not the case then the list of fiducial outcomes $\{\g_1, \g_2, \ldots \}$ would not be complete. This follows from the fact that, operationally, a measurement is any physical process that produces a classical outcome, irrespectively of whether it involves dynamics, interactions with other systems, etc.
Hence, if something is ``observable'' it must produce some effect on the statistics of some measurements.

The set of all vectors \eqref{omega vector} of system $\A$ is denoted by $\Omega_\A$.
Using the fact that the components of $\omega_\lambda$ are probabilities we conclude that, a preparation $\lambda$ consisting of a mixture of preparations $\lambda_x$ with probabilities $p_x$ for $x\in \cal X$ is represented by the vector
\begin{align}\label{mixed omega}
  \omega_{\lambda}
  =
  \sum_{x\in \cal X} p_x\, \omega_{\lambda_x}\ ,
\end{align}
where $\cal X$ can be finite or countably infinite.
Therefore, allowing arbitrary mixtures of preparations makes the set $\Omega_\A$ convex.
The \textbf{pure states} of $\A$ (i.e.~states of maximal knowledge) are the extreme points of $\Omega_\A$.
Also, if the outcome probability $P(\f|\omega_\lambda)$ \replaced{for state \eqref{mixed omega} is independent of whether we know the value of $x$ or not, }{of state \eqref{mixed omega} is independent of whether we forget the the value of $x$ before or after the measurement $\f$} then the function $P(\f|\cdot):\Omega_\A \to [0,1]$ must commute with mixtures,
\begin{align}
  P\!\left( \f \big |\mbox{$\sum_{x}$} p_x \omega_{x} \right)
  =   
  \sum_x p_x P( \f |\omega_{x})\ .
\end{align}
In Appendix 1 of \cite{Hardy_quantum_2001} it is proven that this condition implies that the function $P(\f|\cdot)$ can be 
%extended to the $\real$-linear span of $\Omega_\A$ so that $P(\f|\cdot): \real \Omega_\A \to \real$ is affine. 
%\replaced{This implies that the linear span of the set of OPFs $\real \F_\A$ has dimension equal to that of $\mathbb R\Omega_\A$ plus one; which} 
%{This implies that the set of states $\Omega_\A$ is convex-dual to the set of outcome probability functions $P(\f|\cdot)$, and hence, both have the same dimension.}
\added{linearly extended to the linear span $\mathbb{R}\Omega_\A$ of $\Omega_\A$. Moreover, the states in $\Omega_\A$ separate points in $\mathbb{R}\F_\A$ (namely, no two distinct fiducial outcomes agree on all states), and vice versa (no two distinct states have all outcome probabilities equal), hence $\dim \mathbb{R}\Omega_\A=\dim\mathbb{R}\F_\A$. Furthermore, since $0$ is not in the affine hull of $\Omega_\A$, it follows $\dim\F_\A=\dim\Omega_\A+1$. This generalises the fact that set of Hermitian matrices has dimension equal to that of the density matrices plus one.}

This theory-independent representation of states and measurements is analogous to the density matrices and POVMs of quantum theory.
In particular, if we perform the above construction on a quantum system $\A$ (with quantum measurements) with Hilbert space $\complex^2$ then $\Omega_\A$ is affinely equivalent to the Bloch ball. 
In Section~\ref{sec:assumptions} we adapt this formalism to analyse the family of theories resulting from modifying the measurement postulates of quantum mechanics and leaving the rest of postulates unchanged.
In this case, the set of extreme points of $\Omega_\A$ is the projective space P$\complex^a$, but the mixed states need not correspond to the density matrices.
%The affine dimension of $\Omega_\A$ is equal to the number of parameters that are necessary to specify a mixed state.

Next we enumerate some consequences of the operational framework which are not necessarily valid in Kent's framework \cite{Kent}:
\begin{enumerate}
  \item Each outcome probability function $P(\f|\cdot):\Omega_\A \to [0,1]$ is completely characterised by its action on the pure states of $\Omega_\A$.
  
  \item Each outcome probability $P(\f|\omega)$ is an affine function of the mixed state $\omega$, which is a density operator in quantum theory but not in theories with modified measurement postulates. Therefore, knowing the fiducial probabilities $P(\g_i|\omega)$ allows us to predict the probability of any outcome $P(\f|\omega)$, as in quantum theory. %This invalidates Kent's argument that ``the possibility of state estimation'' assumption does not guarantee the possibility of state estimation.
  
  %\item The relevance of the equivalence classes of ensembles is that they contain all the information about how to assign outcome probabilities to the pure states.
  
  \item The set of preparations $\Lambda_\A$ of a system $\A$ also includes those involving an interaction with another system $\B$.
  Therefore, even if systems $\A$ and $\B$ are in a pure entangled state, the (marginal) $\omega$-state of system $\A$ can be written as a proper mixture of the pure states of $\Omega_\A$. 
  %The reverse is not necessarily true: .  
  That is, the (local) statistics of improper mixtures can be simulated with proper mixtures (the converse is not necessarily true: in some theories there exist mixed states which are not obtainable as the marginal of pure multipartite states~\cite{Galley_impossibility_2018}).   
  
%  \item In \cite{MGM}  The assumption ``possibility of state estimation'' applies to systems
\end{enumerate}

\subsection{No signalling}\label{sec:signalling}

\noindent
In our framework, a \textbf{physical system}  is a set of degrees of freedom on which we can perform preparations and measurements producing outcomes with relative frequencies \replaced{that tend to a definite value if we repeatedly prepare the same physical setup and make the same measurement.}{that tend to the same value every time we prepare the same physical setup.} This property allows us to assign states $\omega$ to preparations, irrespectively of what happens outside our lab.

Instead of this frequentist view, we may alternatively adopt a Bayesian interpretation of probabilities: given a preparation procedure, an agent will hold beliefs about future outcomes of operations on the corresponding physical system $S$, and these beliefs will be unchanged if she learns that other agents are performing operations on other systems that are not currently interacting with $S$.

From this perspective, the only thing that we can safely consider a physical system in a signalling theory is the entire universe. 
In a signalling theory, the statistics of any part of the universe can depend on the measurements performed on other parts, even if we are able to freeze the dynamical interaction between them. This lack of well-defined statistics in terms of preparations prevents the assignation of a state in the sense of \eqref{omega vector}. 

Kent stresses that the violation of ``no signalling'' in his theories is not in contradiction with relativistic causality, since the effect of a measurement on the statistics of separated parts is restricted to the future lightcone of the measurement.
But in any case, Kent's theories involve a notion of physical system and state different than ours. 

%To illustrate this, consider the preparation of a spin-1/2 particle, which produces a small amount of electromagnetic radiation that escapes the laboratory. After this preparation we perform a measurement on the particle, and repeat the procedure many times to record the relative frequency. In a signalling theory, this relative frequency depends on the presence of an observer measuring the electromagnetic radiation outside the lab. On the days where the external observer is present the relative frequency will be different than the days where the observer is not present. Hence, we cannot associate a state to our preparation and measurement.

%HARDY AXIOM 0. Relative frequencies measured by taking the proportion of times a particular outcome is observed) tend to the same value (which we call the probability) for any case where a given measurement is performed on a ensemble of $n$ systems prepared by some given preparation in the limit as $n$ becomes infinite.  

\section{Kent's post-quantum measurement postulates}
\label{sec:Kent theories}

\noindent
In this section we describe and analyse Kent's modifications of quantum mechanics, and show that they are not counter-examples of our non-existence claim in \cite{MGM}.

\subsection{State readout devices and the corresponding structure of pure states}\label{sec:SRD}

\noindent
Kent introduces a hypothetical measurement device $\mathsf{M_{sr}}$ for a quantum system $\A$ which works in the following way. Let $\cal A,B$ be the Hilbert spaces of systems $\A,\B$, where $\B$ can be any environment of $\A$ including the rest of the universe.
When the joint state of $\A\B$ is pure $\ket\psi \in \cal A \otimes B$ then the outcome of $\mathsf{M_{sr}}$ is a classical description of the reduced density matrix $\rho_\A = \tr_{\cal B} \proj\psi$ of system $\A$ in a fixed basis.
When the joint state of $\A\B$ is a proper mixture, like $\ket{\psi_x}\in \cal A \otimes B$ with probability $p_x$ for $x\in \cal X$, then the outcome of $\mathsf{M_{sr}}$ is \added{(a classical description of)} the reduced density matrix $\rho_x = \tr_{\cal B} \proj{\psi_x}$ with probability $p_x$.
%This is very natural with the idea that proper mixtures are not fundamental.

Let us see that when $\ket\psi$ is entangled this measurement device allows for signalling from $\B$ to $\A$. This can be done by performing a projective measurement $\{Q_x\}$ on $\cal B$ which transforms the state $\ket\psi$ into the proper mixture of states $\ket{\psi_x} = (p_x)^{-1/2} (\unity_\A \otimes Q_x)\ket\psi$ with probability $p_x = \bra\psi\unity_\A \otimes Q_x\ket\psi$. 
A clever choice of measurement $\{Q_x\}$ would make the reduced density matrix of $\ket{\psi_x}$ different from that of $\ket\psi$, for all $x$, thereby implementing the signalling.
Kent assumes that the effect on system $\A$ of measuring system $\B$ is restricted to the future lightcone of the measurement, and hence, it is consistent with special relativity.
%We discuss the absence of signalling in our framework in Section~\ref{sec:signalling}.

Now, let us apply the operational framework introduced in Section~\ref{sec:GPT} to the theory resulting from supplementing quantum mechanics with the hypothetical device $\mathsf{M_{sr}}$. The outcome of $\mathsf{M_{sr}}$ on any ray of the Hilbert space $\cal A$ is a unit-rank density matrix, and the same is true for any proper mixture of rays from $\cal A$. However, when systems $\A$ and $\B$ are entangled the outcome of $\mathsf{M_{sr}}$ is a density matrix with rank higher than one, and hence, cannot be simulated by a proper mixture of rays. 
This implies that the pure states of system $\A$ must include elements that are not rays of $\cal A$. This contradicts the ``states postulate'' of quantum mechanics (according to our reading).

More specifically, in this theory, the set of pure states of $\A$ is the set of density matrices acting on $\cal A$, since each of these produces an outcome probability that cannot be written as the mixture of other outcome probabilities.
These new pure states are perfectly distinguishable via the measurement $\mathsf{M_{sr}}$, hence the state space $\Omega_\A$ is an infinite-dimensional classical-probability space (i.e.~a simplex) with pure states labeled by the density matrices on $\cal A$. The mixed states of this theory are not density matrices.
Also, the signalling phenomena of this theory allows for non-classical correlations between $\A$ and $\B$, mimicking quantum entanglement.

\subsection{Other readout devices}

\noindent
Kent also introduces several other types of hypothetical readout devices, and finite-precision versions of them. 
Like the state readout device $\mathsf{M_{sr}}$, all these other devices distinguish between proper mixed states and improper mixed states (i.e.~density matrices obtained as the partial trace of pure bipartite states). Therefore, all of the corresponding theories have pure states other than the rays of the Hilbert space.

Some of these devices (e.g.~stochastic positive-operator devices) also generate sets of pure states consisting of all density matrices. Other devices, like the finite-precision ones, probably have different sets of pure states. But in all of them there are non-quantum pure states.

\subsection{Stochastic positive-operator devices and the post-measurement state update rule}

\noindent
All hypothetical readout devices introduced by Kent have the same post-measurement state update rule: leave the global (and hence also the local) state unchanged. For example, the stochastic positive-operator device $\mathsf{M_{spo}}$ can be configured with a classical description of a POVM $\{X_i\}_{i=1}^n$ for Hilbert space $\cal A$. Then, when applying $\mathsf{M_{spo}}$ on system $\cal A$ prepared in a state $\ket\psi \in \cal A\otimes B$, it produces the outcome $i$ with probability $\bra\psi (X_i \otimes \unity_\B) \ket\psi$, and as mentioned above, the post-measurement state is still $\ket\psi \in \cal A\otimes B$.

Ideally, we can perform the measurement $\mathsf{M_{spo}}$ infinitely many times with different POVM configurations, until we completely learn the reduced density matrix $\rho_\A = \tr_{\cal B} \proj\psi$ of $\A$.
Therefore, the theory resulting from supplementing quantum mechanics with $\mathsf{M_{spo}}$ also contains the measurement $\mathsf{M_{sr}}$. 
Hence, in this theory, the set of pure states is again the set of all density matrices.
This illustrates the fact that when quantum mechanics is supplemented with an extra measuring device $\mathsf M$, then the resulting theory might include other extra measurements beyond $\mathsf M$. 
This observation is relevant to analyse Kent's \added{attempted} refutation of our derivation of the post-measurement state-update rule.

According to the standard quantum postulates, all outcome probability functions are of the form
\begin{align}\label{Q opfs}
  P(\f|\psi) = \bra\psi X_\f \ket\psi ,
\end{align}
where $X_\f$ is the POVM element associated to outcome $\f$.
In \cite{MGM} we prove that any measurement postulate having the quantum probabilistic structure \eqref{Q opfs} must have the quantum post-measurement state-update rule.
Kent \added{attempts to} refute this proof by noting that a $\mathsf{M_{spo}}$ device produces the same probabilities \eqref{Q opfs} than ordinary quantum theory but a different post-measurement state-update rule.
However, within our operational framework, a measurement is any valid process which produces a classical outcome.
In particular, we can consider the joint outcome $(i,j)$ of first, apply measurement $\mathsf{M_{spo}}$ obtaining outcome $i$, and second, apply an ordinary POVM measurement obtaining outcome $j$. In what follows we show that such a composite measurement has a probabilistic structure different than that of quantum theory.

After performing $\mathsf{M_{spo}}$ with input $\{X_i\}_{i=1}^n$ on state $\ket\psi$ we obtain outcome $i$ with probability $P(i)= \bra\psi X_i\ket\psi$ and the post-measurement state $\ket\psi$.
This is followed with a POVM measurement $\{Y_j\}_{j}^m$, which produces outcome $j$ with probability $P(j|i)= \bra\psi Y_j\ket\psi$. (Note that $P(j|i)$ is independent of $i$.)
The probability for the outcome $(j,i)$ of the composite measurement is
\begin{align}
  P(j,i) = P(j|i) P(i) = 
  \bra\psi Y_j\ket\psi\! \bra\psi X_j\ket\psi,   
\end{align} 
which is a non-linear function of $\proj\psi$, therefore this probability is not of quantum type.

In summary, we have invalidated Kent's \replaced{argument against}{refutation of} our derivation of the post-measurement state-update postulate in two ways. First, we have proven that his counter-example is a theory whose set of pure states is not the projective Hilbert space. Second, we have proven that his counter-example \replaced{theories include}{theory includes} measurements with non-quantum statistics.

\subsection{Entropy meters and the dimensionality of mixed states}

\noindent
In this section we analyse the theory resulting from supplementing quantum mechanics with a finite-precision entropy meter $\mathsf{M_{fpem}}$, as defined in \cite{Kent}. We show that, despite the fact that the measurement $\mathsf{M_{fpem}}$ has finitely-many outcomes, in the theory resulting from supplementing quantum mechanics with this measurement the set of mixed states of a $\complex^2$ system has infinite dimension.

Kent defines the post-quantum measurement device $\mathsf{M_{fpem}}$ in the following way. If  two systems $\A, \B$ are described by a pure state $\ket\psi \in \cal A\otimes B$ then feeding system $\A$ into $\mathsf{M_{fpem}}$ produces as outcome a finite-precision numerical value of the Von Neumann entropy 
\begin{align}
  S(\rho_\A) = -\tr(\rho_\A \log_2\rho_\A)  
\end{align}
of the reduced density matrix $\rho_\A = \tr_{\cal B} \proj\psi$.
That is, the outcome of $\mathsf{M_{fpem}}$ is a number 
\begin{align}
  k \in \{0, .01, .02, \ldots, (\log_2 d_{\A}-.01)\}  
\end{align}
satisfying $k \leq S(\rho_\A) < k+.01$, where $d_\A$ is the dimension of $\cal A$.
%For what comes next, it is important to note that the probability of outcome $\mathsf{ M_{fpem}} = k$, denoted $P(\mathsf{ M_{fpem}}\! = k|\rho_\A)$, is not a polynomial function of $\rho_\A$. This is also true if consider the 2-Renyi entropy $S(\rho_\A) = -\log_2\tr\rho^2_\A$ instead of the Von Neumann one.

Following our operational framework, the new theory with measurement $\mathsf{M_{fpem}}$ must also include the  measurement $\mathsf N$ for qubit ${\cal A} = \complex^2$ defined as follows.
%(This construction is based on a particular case of property (9) from \cite{MGM}.)
For any state $\ket\varphi = \alpha\ket0 +\beta\ket1 \in \cal A$ apply the following steps:
\begin{enumerate}
  %\item apply $U$ to $\ket\varphi$
  
  \item prepare qubit $\B$ with $\ket0 \in \cal B$, 
  
  \item apply the CNOT gate to the two qubits $\cal A\otimes B$ producing the state
  \begin{align}
    \alpha\ket0_\A \ket0_\B+
    \beta\ket1_\B \ket1_\B
  \end{align}
  
  \item perform $\mathsf M_\mathsf{fpem}$ on qubit $\A$, which produces an outcome $k$ being an approximation to the number
  \begin{align}
    S(\rho_\A) = 
    - |\alpha|^2 \log_2 |\alpha|^2
    -|\beta|^2 \log_2 |\beta|^2.
  \end{align}
\end{enumerate}
This procedure results in an outcome probability of $P\big(\mathsf N \!=\! k \big| \alpha \ket0 \!+\! \beta\ket1 \big) =1$ if 
\begin{align}\label{OPF 3 step}
  k\leq - |\alpha|^2 \!\log_2 \! |\alpha|^2 -\! |\beta|^2 \!\log_2 \! |\beta|^2 <k\! +\! .01,   
\end{align}
and $P\big(\mathsf N \!=\! k \big| \alpha \ket0 \!+\! \beta\ket1 \big) =0$ otherwise.
The fact that this outcome probability function is not a polynomial in $\alpha, \beta$ and their complex conjugates $\bar\alpha, \bar\beta$ implies that the family of functions $P\big(\mathsf N \!=\! k \big| U\ket\varphi \big)$ for all $U\in $ SU(2) generates an infinite-dimensional linear space of functions P$\complex^2\to \mathbb R$, which is a subspace of the linear space $\mathbb R \F_\A$ generated by all outcome probability functions of the theory. By recalling that the linear space $\mathbb R \F_\A$ is dual to the linear space spanned by the mixed states $\Omega_\A$, we conclude that $\mathbb R\Omega_\A$ is infinite-dimensional, too.

If instead of the Von Neumann entropy we use the second-order R\'enyi entropy then the logarithms in \eqref{OPF 3 step} are replaced by polynomials. However, the function $P\big(\mathsf N \!=\! k \big| \alpha \ket0 \! +\! \beta\ket1 \big)$ with variables $\alpha, \beta \in \complex$ is still non-polynomial, so the above argument still holds.

\section{Framework and assumptions in \cite{MGM}}
\label{sec:assumptions}

\noindent
In this section we adapt the operational framework introduced in Section~\ref{sec:GPT} to theories resulting from modifying the measurement postulates of quantum mechanics and leaving the rest of postulates unchanged.
In this case it is more convenient to work with sets of outcomes $\F_\A$ than with state spaces $\Omega_\A$, \replaced{but both methods are equivalent.}
{In any case, $\F_\A$ is the convex-dual of $\Omega_\A$, so the two methods are equivalent.}
We also discuss the ``finite-dimensionality of mixed states'' assumption.

\subsection{The OPF formalism}

\noindent
%the case where the pure states of a system $\A$ correspond to rays in a Hilbert space $\psi \in {\rm P}\complex^a$ it is convenient
According to the standard quantum measurement postulates, each outcome $\f \in \F_a$ of system $\complex^a$ is mathematically represented by a POVM element $0\leq X_\f \leq \unity$.
If we modify these postulates then another representation may arise.
But we would like a representation that is valid in any theory satisfying the non-measurement quantum postulates (i.e.~states, transformations and composite systems postulates). 
In this general situation, a convenient way to mathematically represent each outcome $\f \in \F_a$ of system a $\complex^a$ is by its corresponding probability function $\f: {\rm P}\complex^a \to [0,1]$ satisfying $\f(\psi) = P(\f|\psi)$ for all $\psi\in {\rm P}\complex^a$. Then we say that each element $\f: {\rm P}\complex^a \to [0,1]$ of $\F_a$ is an outcome probability function or OPF.
A measurement with $n$ outcomes is represented by $n$ OPFs $\f_1, \ldots, \f_n \in \F_a$ satisfying the normalisation condition
\begin{align}\label{cond:normalisation}
  \f_1 (\psi) +\cdots +\f_n(\psi) = 1 \ ,
\end{align}
for all $\psi\in$ P$\complex^a$.

In order to specify an alternative measurement postulate (in the context of the other quantum postulates) we need to provide an OPF set $\F_a$ for every Hilbert space $\complex^a$, with $a=2,3,4,\ldots, \infty$. (We denote by $\complex^\infty$ the Hilbert space of countably-infinite dimension.)
But this is not enough.
%In order to fully specify an alternative measurement postulate we also need to detail how two effects $\f \in \F_a$ and $\mathbf h \in \F_b$ of systems $\complex^a$ and $\complex^b$ are represented in the set of effects of the composite system $\complex^a \otimes \complex^b \cong \complex^{ab}$.This information is contained in the star-product $\f \star \mathbf h \in \F_{ab}$, which 
%
In order to fully specify an alternative measurement postulate we also need to provide the joint probability $P(\f,\mathbf h|\psi)$ of any pair of outcomes $\f \in \F_a$ and $\mathbf h \in \F_b$ of systems $\complex^a$ and $\complex^b$ on any pure state $\psi$ of the composite system $\complex^a \otimes \complex^b \cong \complex^{ab}$.
Algebraically, this information is captured by a product of OPFs (analogous to the quantum tensor-product between POVM elements). The $\star$-product is defined as the family of maps $\star: \F_a \times \F_b \to \F_{ab}$ for all pairs of dimensions $(a,b)$ satisfying
\begin{align}
  P(\f,\mathbf h|\psi) = (\f\star \mathbf h)(\psi)\ .
\end{align}
In particular, the $\star$-product specifies a map from any pair of outcomes $\f \in \F_a$ and $\mathbf h \in \F_b$ to its joint outcome $(\f\star \mathbf h) \in \F_{ab}$ for the composite system $\complex^a \otimes \complex^b$.

In algebraic terms, an alternative measurement postulate for quantum mechanics is:
\begin{enumerate}
  \item a family of OPF sets $\F_a$,
  \item and a product $\star: \F_a \times \F_b \to \F_{ab}$,
\end{enumerate}  
for all $a,b\in \{2,\ldots, \infty\}$, satisfying the consistency constraints (\ref{prop1}, \ref{prop2}, \ref{prop3}, \ref{prop4}, \ref{prop5}, \ref{prop6}, \ref{prop7}, \ref{prop8}, \ref{prop9}, \ref{prop10}) detailed in the following two subsections.
(All these constraints are written below inside boxes.)
An alternative measurement postulate could also provide additional structure, like which outcomes $\f\in \F_a$ constitute a measurement. But the minimum content of a measurement postulate is a pair $(\{\F_a\}_{a=2}^\infty, \star)$.

%In most theories, the $\omega$-vector representation of states is redundant, because some outcome probabilities $P(\tilde \f|\lambda)$ are a function of other outcome probabilities $P(\f^i|\lambda)$ for $i\in \cal I$. In this case there is an affine map between the space of $\omega$-vectors and another space with lower dimension. For the case of a qubit $\complex^2$

\subsection{Properties of the OPF sets $\F_a$}

\noindent
As mentioned in the introduction, in the operational framework, the composition of a measurement with a unitary transformation produces another measurement of the theory.
Therefore, each OPF set $\F_a$ is closed under composition with unitaries
\begin{align}\label{prop1}
  \boxed{\f \circ U \in \F_a}
\end{align}
for any $\f \in \F_a$ and any $U\in {\rm SU}(a)$.
Similarly, the mixture of measurements produces another another valid measurement, hence each OPF set $\F_a$ is convex
\begin{align}\label{prop2}
  \boxed{\sum_x p_x\, \f^x \in \F_a}
\end{align}
for any $\f^x\in \F_a$ and any probability distribution $p_x$.

A measurement with a single outcome has OPF $\u_a\in \F_a$ satisfying $\u_a (\psi)=1$ for all $\psi \in {\rm P}\complex^a$, due to the normalisation of probabilities. Since an experimentalist can perform this trivial measurement in any situation we assume that this ``unit OPF'' is included
\begin{align}\label{prop3}
  \boxed{\u_a \in \F_a}\ .
\end{align}
The unit OPF also allows us to write condition \eqref{cond:normalisation} as $\f_1 +\cdots +\f_n = \u_a$. 

We can also consider a measurement that produces a random outcome with probability distribution $(p,1-p)$ independently of the state of the system $\psi\in {\rm P}\complex^a$.
In the limit $p\to 0$ the first outcome has probability 0 for all $\psi \in$ P$\complex^a$. 
\replaced{Motivated by this, it is useful}{
Since this is a valid procedure we also need} to include the zero OPF 
\begin{align}\label{prop4}
  \boxed{\mathbf 0_a \in \F_a}  
\end{align}
defined by $\mathbf 0_a(\psi)=0$ for all $\psi \in$ P$\complex^a$. 

\added{Next, consider the following process: (i) sample the random variable $x$ with probability $p_x$, (ii) transform the system with $U\in \SU(a)$ and (iii) perform measurement $\f_1^x, \f_2^x, \ldots \in \F_a$. 
If the average probability of outcome $k$ is independent of whether we know the value of $x$ or not then
\begin{align}
  \sum_x p_x \left(\f_k^x \circ U\right)
  = \Big(\sum_x p_x \f_k^x \Big) \circ U\ ,
\end{align}
which implies that the action $U: \real\F_a \to \real\F_a$ is affine. Additionally, by definition of the zero OPF we have 
\begin{align}
  \mathbf 0_a \circ U = \mathbf 0_a\ ,  
\end{align}
which implies that the action is not only affine, but linear. In summary, the action $U: \real\F_a \to \real\F_a$ is a representation of the group $\SU(a)$ on the vector space $\real\F_a$.}

\added{Finally, note that} we can implement a measurement on system $\complex^a$ by first, preparing an ancilla in the state $\varphi \in {\rm P}\complex^b$ and second, perform the joint measurement $\f \in \F_{ab}$ on the composite system $\complex^a \otimes \complex^b$.
Therefore, the following must be a valid OPF:
\begin{align}\label{prop5}
  \boxed{\f(\cdot \otimes\varphi) \in \F_a} 
\end{align}
for any $\f \in \F_{ab}$ and any $\varphi \in {\rm P}\complex^b$.

\subsection{Alternative measurement postulates for single systems} 

\noindent
In this subsection we present the characterisation of all finite-dimensional OPF sets $\F_a$ which satisfy assumptions \eqref{prop1} and \eqref{prop2}. Since in this subsection we are only concerned with single systems, we ignore the $\star$-product.
This result was proven in \cite{Galley_classification_2017} using the representation theory of $\SU(d)$, and it illustrates that, if we ignore the consistency constraints arising in composite systems then there are infinitely-many alternatives to the measurement postulates. 
Interestingly, some of these alternatives are also consistent on bi-partite systems (see Section 4 of \cite{Galley_impossibility_2018}).

For a fixed Hilbert space dimension $a$, every finite-dimensional set $\F_a$ is characterised by a positive integer $k$, and for every function $\f\in \F_a$ there is an Hermitian matrix $F$ acting on the $k$-fold symmetric subspace of $(\complex^a)^{\otimes k}$ such that
\begin{align}\label{alter BR}
  \f(\psi) 
  = \tr\!\left(F \ket\psi\!\! \bra\psi^{\otimes k} \right)\ .
\end{align}
In addition to specifying the parameter $k$, an alternative measurement postulate may involve additional constraints on the set of valid $F$-matrices. The case $k=1$ corresponds to standard quantum mechanics.
In 1974, Bogdan Mielnik argued in \cite{Mielnik_generalized_1974} that the family of OPFs \eqref{alter BR} form consistent modifications to the quantum measurement postulates for single systems, however he did not prove that all consistent alternative measurement postulates are in this family.

\subsection{Properties of the $\star$-product and no-signalling}
\label{sec:prop star}

\noindent
The option of describing the system $\complex^a$ as part of a larger system $\complex^a \otimes \complex^b$ is a subjective choice which must not affect the predictions of the theory, therefore, the embedding of $\F_a$ into $\F_{ab}$ provided by the $\star$-product must preserve the structure of $\F_a$. In particular, we impose that the mixing structure is preserved 
\begin{align}\label{prop6}
  \boxed{
  \left(\mbox{$\sum_x p_x$} \f^x \right) \star \mathbf h = 
  \mbox{$\sum_x p_x$} \left(\f^x \star\mathbf h \right)},
\end{align}
and the SU($a$) action is preserved too
\begin{align}\label{prop7}
  \boxed{
  (\f \circ U)\star \mathbf h 
  = (\f\star \mathbf h)\circ (U\otimes \unity_b)}\ .
\end{align}
And likewise for the other party $\F_b$. 

The $\star$-product must also preserve the normalisation of probabilities. That is, if the measurements $\{\f_i\} \subseteq \F_a$ and $\{\mathbf h_j\} \subseteq \F_b$ satisfy the normalisation condition \eqref{cond:normalisation} then 
\begin{equation}
    \label{cond:u*u=u}
    \sum_{i,j} (\f_i \star \mathbf h_j)(\psi) = 1\ ,
\end{equation}
for all rays $\psi$ in $\complex^a \otimes \complex^b$.
This condition is equivalent to 
\begin{align}\label{prop9}
  \boxed{\u_a \star \u_b = \u_{ab}}\ .
\end{align}

We also require that the event consisting of jointly observing the zero-probability outcome $\mathbf 0_a$ on system  $\complex^a$ and the unity-probability outcome $\u_b$ on $\complex^b$ has zero probability $(\mathbf 0_a \star \u_b)(\psi)=0$ on all rays $\psi$ in $\complex^a \otimes \complex^b$. This is equivalent to 
\begin{align}\label{prop8}
  \boxed{
  \mathbf 0_a \star \u_b = \mathbf 0_{ab}
  }\ .
\end{align}

In Lemma~2 of \cite{MGM} it is proven that conditions \eqref{prop6} and \eqref{prop8} imply that the action of the $\star$-product on $\F_a \!\times\! \F_b$ can be extended to an action on $\mathbb R \F_a \!\times \mathbb R \F_b$ that is linear in each argument, where $\mathbb R \F_a$ denotes the space of linear combinations with real coefficients of all the functions in $\F_a$.

This bi-linearity implies the no-signalling condition: the marginal probability of outcome $\g\in \F_b$ on system $\complex^b$ is independent on whether we measure observable $\{\f_i\}\in \F_a$ or observable $\{\mathbf h_j\}\in \F_a$ on system $\complex^a$,
\begin{align}
  \nonumber
  &\sum_i [\f_i\star \g](\psi)
  = [(\mbox{$\sum_i$} \f_i)\star \g](\psi)
  = [\u_a\star \g](\psi)
  \\ &\hspace{6mm} = 
  [(\mbox{$\sum_j$} \mathbf h_j)\star \g](\psi)
  = \sum_j [\mathbf h_j\star \g](\psi)\ ,
\end{align}
for all rays $\psi$ in $\complex^a \otimes \complex^b$.
This also tells us that the reduced state on system $\complex^b$ of any bipartite pure state $\psi\in {\rm P}(\complex^a \otimes \complex^b)$ is given by the map $\F_b \to [0,1]$ defined as $\g \mapsto (\u_a \star \g)(\psi)$.

Consistency with the compositional structure of quantum mechanics requires that the reduced state of $\psi \otimes \varphi \in \complex^a \otimes \complex^b$ on $\complex^a$ is $\psi$, or equivalently,
\begin{align}\label{prop9} 
  \boxed{
  (\f\star\u_b)(\psi\otimes\varphi)
  =\f(\psi)
  }
\end{align}
for all $\psi\in\complex^a$, $\varphi\in\complex^b$ and $\f\in\F_a$.
This implies (Lemma~13 in \cite{MGM}) that product pure states have no correlations 
\begin{align}
  (\f\star\g)(\psi\otimes\varphi)
  =\f(\psi)\, \g(\varphi)\ .
\end{align}

Finally, we require that choosing to describe the tripartite system $\complex^a \otimes \complex^b \otimes \complex^c$ as the bipartite system $\complex^a \otimes (\complex^b \otimes \complex^c)$ or the bipartite system $(\complex^a \otimes \complex^b) \otimes \complex^c$ does not alter the predictions of the theory.
Therefore, the $\star$-product must be associative
\begin{align}\label{prop10} 
  \boxed{
  \f\star(\g\star\mathbf h) 
  = (\f \star\g)\star \mathbf h
  }
\end{align}
for any $\f\in \F_a$, $\g\in \F_b$ and $\mathbf h \in \F_c$.
Interestingly, Section 4 of \cite{Galley_impossibility_2018} contains an alternative measurement postulate for quantum mechanics which satisfies all consistency constraints except for associativity \eqref{prop10}. 
%(\ref{prop1}, \ref{prop2}, \ref{prop3}, \ref{prop4}, \ref{prop5}, \ref{prop6}, \ref{prop7}, \ref{prop8}, \ref{prop9}). 
This simple alternative theory is consistent as long as we only represent single and bi-partite systems. 
%Moreover in \TG{CITE} we provide a classification of all consistent single systems with modified measurement postulates and in \TG{CITE} we provide an example of a consistent bi-partite system with modified measurement postulate.

%In order to prove that the only measurement postulates compatible with the rest of quantum postulates is the standard measurement postulate, we need to formalise the notion of ``measurement postulate'' for any theory where pure states are rays in a Hilbert space $\psi\in {\rm P}\complex^d$, transformations include unitaries $U\in$ SU($d$), and the Hilbert space of a composite system is the tensor product.

%The set of mixed states of a system with finite-dimensional Hilbert space is finite-dimensional.

\subsection{The possibility of state estimation}

On top of the operational framework presented above we assume that, if the dimension of $\complex^a$ is finite then the dimension of the linear space $\mathbb R \F_a$ is finite too:
\begin{align}
  \boxed{{\rm dim}\complex^a<\infty
  \ \Rightarrow\ 
  {\rm dim}\mathbb R \F_a<\infty}\ .
\end{align} 
\replaced
{Recall that ${\rm dim} \Omega_a = {\rm dim}\mathbb R \F_a -1$, which is the number of parameters that are required to specify a general mixed state.} 
{Recall that $\mathbb R \F_a$ is convex-dual to $\Omega_a$, and that the dimension of $\Omega_a$ is equal to the number of parameters that are required to specify a general mixed state.} Therefore, this dimensionality assumption is equivalent to the statement that: \emph{each mixed state of a finite-dimensional Hilbert space is characterised by finitely-many parameters.}

At this moment we do not know of a precise connection between the finite-dimensionality of the mixed states of a system $\complex^a$ and the possibility of performing state estimation on $\complex^a$ with finite means. For this reason, in this note we refer to this assumption as ``the finite-dimensionality of mixed states'', with the understanding that it only applies to finite-dimensional Hilbert spaces.  Clearly, infinite-dimensional Hilbert spaces are allowed to have infinite-dimensional sets of mixed states. 

Kent argues that the finite-dimensionality of mixed states is neither necessary nor sufficient for state estimation. We disagree with his claim that the finite dimensionality of mixed states is not sufficient for state estimation. We do agree however with his conclusion that it is not necessary, however we disagree with his argument. In the following two paragraphs we discuss our two claims.

Kent argues that, in a general theory, the relation between the fiducial probabilities \eqref{omega vector} and the rest of the outcome probabilities $P(\f|\omega)$ is arbitrary. Therefore, an exponentially fast estimation of the fiducial probabilities \eqref{omega vector} via observed relative frequencies, could translate to an intractably slow convergence for the accurate prediction of the rest of probabilities $P(\f|\omega)$. We have shown at the end of Section~\ref{sec:GPT} that any function $P(\f|\omega)$ is affine in the fiducial probabilities $\omega$. Therefore, Kent's mismatch of accuracies cannot happen.

Kent also argues that when quantum mechanics is supplemented with readout devices, the predictability of probabilities on ensembles of Hilbert-space rays is not sufficient to predict outcome probabilities on any improper mixed state\deleted{s}. This is correct, but this theory violates the pure state postulate of quantum mechanics. If we had to express the condition of finite-dimensionality of mixed states for Kent's theory, we would require ensembles of any pure state of the theory, which in this case includes not only Hilbert-space rays but also density matrices. %marginals of rays in a bipartite Hilbert space (i.e. density operators). An extension of the OPF framework to theories with different pure states than quantum theory can be found in \TG{CITE} where the condition of finite dimensionality of mixed state in the general case is given in section 2.4.2. 

\section{Closing remarks}

\noindent
The ``finite-dimensionality of mixed states for systems with finite-dimensional Hilbert spaces'' is not a consequence of the operational framework. 
%Also, we agree with Kent on that its relation to the possibility of state estimation is not clear. However, its physical meaning is straightforward, since the dimension of the set of mixed states is equal to the number of parameters that are needed to specify a general mixed state.
Hence, we would like to understand whether this assumption is necessary to single out the quantum measurement postulates, as in \cite{MGM}.
Therefore, we conclude with the following question.
Can we find an alternative to the measurement postulates consistent with the rest of postulates of quantum mechanics (i.e.~satisfying conditions (\ref{prop1}, \ref{prop2}, \ref{prop3}, \ref{prop4}, \ref{prop5}, \ref{prop6}, \ref{prop7}, \ref{prop8}, \ref{prop9}, \ref{prop10})), which does produce a theory with infinite-dimensional sets of mixed states associated to finite-dimensional Hilbert spaces?

\bibliographystyle{unsrtnat}
\bibliography{biblio}

\begin{thebibliography}{10}
\providecommand{\natexlab}[1]{#1}
\providecommand{\url}[1]{\texttt{#1}}
\expandafter\ifx\csname urlstyle\endcsname\relax
  \providecommand{\doi}[1]{doi: #1}\else
  \providecommand{\doi}{doi: \begingroup \urlstyle{rm}\Url}\fi

\bibitem[Kent(2023)]{Kent}
Adrian Kent.
\newblock The measurement postulates of quantum mechanics are not redundant.
\newblock \emph{ArXiv
  \href{https://doi.org/10.48550/arXiv.2307.06191}{arXiv:2307.06191.}}, 2023.

\bibitem[Masanes et~al.(2019)Masanes, Galley, and M\"uller]{MGM}
Llu{\'{\i}}s Masanes, Thomas~D. Galley, and Markus~P. M\"uller.
\newblock The measurement postulates of quantum mechanics are operationally
  redundant.
\newblock \emph{Nature Communications}, 10\penalty0 (1), mar 2019.
\newblock \doi{10.1038/s41467-019-09348-x}.

\bibitem[Hardy(2001)]{Hardy_quantum_2001}
L.~Hardy.
\newblock Quantum theory from five reasonable axioms.
\newblock \emph{ArXiv
  \href{https://doi.org/10.48550/arXiv.quant-ph/0101012}{arXiv:quant-ph/0101012.}},
  January 2001.

\bibitem[Barrett(2007)]{Barrett_information_2005}
Jonathan Barrett.
\newblock Information processing in generalized probabilistic theories.
\newblock \emph{Phys. Rev. A}, 75:\penalty0 032304, March 2007.
\newblock \doi{10.1103/PhysRevA.75.032304}.

\bibitem[Daki\'c and Brukner(2011)]{Dakic_quantum_2009}
B.~Daki\'c and C.~Brukner.
\newblock Quantum theory and beyond: Is entanglement special?
\newblock In H.~Halvorson, editor, \emph{Deep Beauty - Understanding the
  Quantum World through Mathematical Innovation}, pages 365--392. American
  Institute of Physics, 2011.
\newblock \doi{10.1017/CBO9780511976971.011}.

\bibitem[Masanes and M\"uller(2011)]{Masanes_derivation_2011}
Llu{\'\i}s Masanes and Markus~P. M\"uller.
\newblock A derivation of quantum theory from physical requirements.
\newblock \emph{New Journal of Physics}, 13\penalty0 (6):\penalty0 063001,
  2011.
\newblock \doi{10.1088/1367-2630/13/6/063001}.

\bibitem[Masanes et~al.()Masanes, M\"uller, Augusiak, and
  Perez-Garcia]{Masanes_existence_2013}
L.~Masanes, M.~P. M\"uller, R.~Augusiak, and D.~Perez-Garcia.
\newblock Existence of an information unit as a postulate of quantum theory.
\newblock \emph{Proceedings of the National Academy of Sciences}, 110\penalty0
  (41):\penalty0 16373--16377.
\newblock ISSN 0027-8424, 1091-6490.
\newblock \doi{10.1073/pnas.1304884110}.

\bibitem[Galley and Masanes(2018)]{Galley_impossibility_2018}
Thomas~D. Galley and Lluis Masanes.
\newblock Any modification of the {B}orn rule leads to a violation of the
  purification and local tomography principles.
\newblock \emph{{Quantum}}, 2:\penalty0 104, November 2018.
\newblock ISSN 2521-327X.
\newblock \doi{10.22331/q-2018-11-06-104}.

\bibitem[Galley and Masanes(2017)]{Galley_classification_2017}
Thomas~D. Galley and Lluis Masanes.
\newblock Classification of all alternatives to the {B}orn rule in terms of
  informational properties.
\newblock \emph{{Quantum}}, 1:\penalty0 15, July 2017.
\newblock ISSN 2521-327X.
\newblock \doi{10.22331/q-2017-07-14-15}.

\bibitem[Mielnik(1974)]{Mielnik_generalized_1974}
Bogdan Mielnik.
\newblock Generalized quantum mechanics.
\newblock \emph{Communications in Mathematical Physics}, 37\penalty0
  (3):\penalty0 221--256, 1974.
\newblock \doi{https://doi.org/10.1007/BF01646346}.

\end{thebibliography}

% \begin{thebibliography}{99}
% \bibitem{Kent} Adrian Kent, \href{https://doi.org/10.48550/arXiv.2307.06191}{arXiv:2307.06191.}

% \bibitem{MGM} Llu\'is Masanes, Thomas D.~Galley and Markus P.~M\"uller, \href{https://doi.org/10.1038/s41467-019-09348-x}{Nature Communications 10, 1361 (2019).}

% \bibitem{Hardy_quantum_2001} Lucien Hardy, \href{https://doi.org/10.48550/arXiv.quant-ph/0101012}{arXiv:quant-ph/0101012.}

% \bibitem{Barrett_information_2005} Jonathan Barrett, \href{https://doi.org/10.1103/PhysRevA.75.032304?_gl=1*17pozmo*_gcl_au*MTQ0OTg0NTQ3MC4xNzM1NDcxODc4*_ga*ODIxMDY4MzM3LjE3MzU0NzE4Nzc.*_ga_ZS5V2B2DR1*MTczNTQ3MTg3Ny4xLjAuMTczNTQ3MTg3Ny42MC4wLjEwMjU0NDk5NzM.}{Phys.~Rev.~A, vol.~75, p.~032304, 2007.}

% \bibitem{Dakic_quantum_2009} Borivoje  Daki\'c and Caslav Brukner, \href{https://doi.org/10.48550/arXiv.0911.0695}{Deep Beauty--Understanding the Quantum World through Mathematical Innovation (H. Halvorson, ed.), pp.~365–392, American Institute of Physics, 2011.}

% \bibitem{Masanes_derivation_2011} Llu\'is Masanes and Markus P.~M\"uller, \href{https://dx.doi.org/10.1088/1367-2630/13/6/063001}{New Journal of Physics, vol.~13, no.~6, p.~063001, 2011.}

% \bibitem{Masanes_existence_2013} Llu\'is Masanes, Markus P.~M\"uller, Remigiusz Augusiak and David P\'erez-Garc\'ia, \href{https://doi.org/10.1073/pnas.1304884110}{Proceedings of the National Academy of Sciences, vol.~110, no.~41, pp.~16373–16377, 2013.}

% \bibitem{Galley_impossibility_2018} Thomas D.~Galley and Llu\'is Masanes, \href{https://doi.org/10.22331/q-2018-11-06-104}{Quantum, vol.~2, p.~104, 2018.}

% \bibitem{Galley_classification_2017} Thomas D.~Galley and Llu\'is Masanes, \href{https://doi.org/10.22331/q-2017-07-14-15}{Quantum, vol.~1, p.~15, 2017.}

% \bibitem{Mielnik_generalized_1974} Bogdan Mielnik, \href{https://doi.org/10.1007/BF01646346}{Communications in Mathematical Physics, vol.~37, no.~3, pp.~221–256, 1974.}

% \end{thebibliography}
% %\bibliography{refs}{}

\end{document}